\documentclass[oneside,12pt]{article}
\usepackage[a4paper, total={6.5in, 10in}]{geometry}
\usepackage[UKenglish]{babel}
\usepackage{lmodern} 
\usepackage{float}
\usepackage{caption}
\usepackage{subcaption}
\usepackage{amsmath}
\usepackage{amssymb}
\usepackage{amsthm}
\usepackage{amsfonts}
\usepackage{multirow}
\usepackage{graphicx}
\usepackage{hyperref}
\usepackage{color,soul}
\usepackage{cite}

\usepackage{lipsum}

\let\OLDthebibliography\thebibliography
\renewcommand\thebibliography[1]{
  \OLDthebibliography{#1}
  \setlength{\parskip}{0pt}
  \setlength{\itemsep}{0pt plus 0.3ex}
}

\begin{document}

\title{\bf Resolving ECRH deposition broadening due to edge turbulence in DIII-D by 3D full-wave simulations}
\author{M.B. Thomas$^1$, M.W. Brookman$^{2}$, M.E. Austin$^{2}$, \\ A. K{\"o}hn$^3$, R.J. La Haye$^4$, J.B. Leddy$^1$, R.G.L. Vann$^1$ \\ and Z. Yan$^5$ \\
\\\footnotesize$^1$ York Plasma Institute, Department of Physics, University of York, Heslington, York, YO10 5DD UK
\\\footnotesize$^2$ Institute for Fusion Studies, University of Texas at Austin, TX 78712, USA
\\\footnotesize$^3$ Max Planck Institute for Plasma Physics, D-85748 Garching, Germany
\\\footnotesize$^4$ General Atomics, PO Box 85608, San Diego, CA 92186-5608, USA
\\\footnotesize$^5$ University of Wisconsin-Madison, Madison, WI, 53706, USA
\\
\\
\footnotesize Email: matt.thomas@york.ac.uk}

\date{\today}
\maketitle

\begin{abstract}
Edge plasma density fluctuations are shown to have a significant effect on the electron cyclotron resonance heating (ECRH) beam in the DIII-D tokamak. This work discusses the theoretical modelling portion of an effort to understand ECRH scattering by turbulence. The corresponding experimental work can be found in M.W. Brookman \textit{et. al.}:\ \textit{Submitted to Nuclear Fusion} (2017). Experimental measurements of the ECRH deposition profile have been taken in three operating scenarios: L-mode, H-mode and negative triangularity. Each scenario corresponds to distinct turbulence characteristics in the edge region through which the beam must propagate. The measured ECRH deposition profile is significantly broadened by comparison to the profile predicted by the ray tracing code TORAY-GA and has been shown to scale with the severity of edge turbulence. Conventional ray tracing does not include the effects of turbulence and therefore a 3D full-wave cold plasma finite difference time domain code EMIT-3D is presented and used for the simulations. The turbulence is generated through the Hermes model in the BOUT++ framework which takes as input the measured time averaged electron density, temperature and magnetic field profiles for the specific shot in question. The simulated turbulence is constrained to match the experimentally measured (by use of the BES and DBS systems) correlation length and normalised fluctuation levels. The predictions of the beam broadening from the simulations are found to agree very well with the experimentally-observed broadening in all cases: L-mode, H-mode and negative triangularity. Due to the large gradients within the H-mode edge, the resolution uncertainty and error in the measurement from Thomson scattering and BES diagnostics result in a spread in the simulated turbulence amplitude. In light of this a parameter scan through the range in experimental diagnostic measurement uncertainty has been conducted to explore the impact on beam broadening predictions. 

\end{abstract}

\newpage

\section{Introduction}
\label{Sec:Intro}

Many current and future magnetically confined fusion devices will rely on collimated electromagnetic beams for precision heating and current drive in the plasma. ITER will use O-mode beams to provide electron cyclotron resonance heating (ECRH) and current drive (ECCD) from an upper port launcher. One primary objective is to stabilise the expected $m = 3,~n = 2$ and the $m = 2,~n = 1$ neoclassical tearing modes (NTMs) \cite{Henderson2008, Zohm2005}. These magneto-hydrodynamic (MHD) instabilities are magnetic islands that develop on rational $q$ surfaces flattening the pressure gradient across the NTM resulting in radially increased particle and energy transport. The island is driven unstable by loss of bootstrap current due the flattened pressure profile which causes the island to grow \cite{Hegna1997}. As the island grows it generates eddy currents in the wall of the vessel causing the plasma rotation to slow and eventually lock, resulting in a disruption where the stored energy of the plasma is dumped to the vessel wall \cite{Brand2012, Zohm2005}. This instability is therefore dangerous in particular for ITER due to the large increase in stored energy of the plasma as compared with current devices. Experimental success has been found in mitigating and suppressing NTMs by use of ECCD in ASDEX-U \cite{Gantenbein2000, Zohm2007}, DIII-D \cite{Petty2004, LaHaye2002} and JT-60U \cite{Isayama2003}.

\vspace{5mm}

The stabilising ECCD is effective within the O-point of the NTM. Consequently if the deposition profile is broadened larger than the island width or the alignment of the beam is not centred on the O-point then the stabilising efficiency is reduced \cite{Petty2004}. Accurate targeting of the NTM can be achieved through active control diagnostics such as the 1 cm accuracy achieved by the DIII-D Plasma Control System (for reference a saturated $3/2$ island width in DIII-D is $4-7$ cm) \cite{LaHaye2002, Humphreys2006}. However it has been widely observed that significant broadening of the ECRH deposition profile occurs to a varying degree ($1.4 - 2.7$ times the width calculated by a beam tracer) across a range of devices such as ASDEX-U \cite{Kirov2002}, DIII-D \cite{Petty2002, Brookman2017, Brookman2017b} and TCV \cite{Decker2012} in a range of operating scenarios. This is of concern for ITER as it raises the power requirement to suppress a given island. Previous work suggests that for the current ITER design full ECCD power from all gyrotrons is needed to suppress the 3/2 island if the deposition profile is broadened by $> 200\%$ compared to ray tracing calculations \cite{LaHaye2001}. This means that there will not be any power left to stabilise the 2/1 island. More recent work suggests that a deposition broadening by $> 100\%$ may necessitate power modulation for NTM suppression \cite{Poli2015}. Most important however, is the performance Q value of ITER will be reduced if it becomes necessary to increase the power requirements for NTM suppression or to continuously use ECCD power. This is of direct consequence for one of ITER's main goals.

\vspace{5mm}

The broadening of ECRH deposition profiles has usually been attributed to the radial transport of fast electrons through diffusion \cite{Kirov2002, Petty2002, Bertelli2009}. Though this is able to explain the broadening profile observed, the necessary bremsstrahlung emission that would be associated with the large radial transport of fast electrons is not observed in experiment \cite{Nikkola2003, Decker2011}. Furthermore it is noted in Ref. \cite{Poli2015} that if the profile is broadened by scattering of the beam by density fluctuations in the edge then the effect of finite transport is reduced. With a view to ITER, direct simulation of the 3/2 and 2/1 surfaces by the gyrokinetics code GKW \cite{Peeters2009} constrains the diffusion coefficient to a value that would account for broadening of the deposition profile of only 10\% \cite{Casson2015}. Indeed recent experimental \cite{Brookman2017, Decker2012} works and the corresponding experimental work to this paper \cite{Brookman2017b} have shown a significant broadening effect which is attributed to scattering of the ECRH beam by density perturbations in the edge.

\vspace{5mm}

The most common treatment of wave propagation is through the use of ray and beam tracing codes. Ray tracing tracks the path of a single ray through a plasma with slowly varying refractive index, useful for predicting the beam path through a tokamak plasma. Beam tracing takes this approach a step further by launching many rays and by keeping track of the phase of each ray, can account for the interference between the rays to reconstruct a beam profile at any desired location. An extensive comparison of a selection of widely used codes is presented in Ref. \cite{Prater2008}. These codes do not incorporate the scattering effects of turbulence on the beam propagation. However, due to the necessity to predict beam broadening in ITER for considering the power required to stabilise NTMs, attempts to include statistical models of turbulence have been undertaken. A number of different methods exist which have been able to produce predictions of significant broadening of the ECRH beam on ITER in some scenarios of up to 100\% \cite{Tsironis2009, Bertelli2010, Sysoeva2015, Peysson2011, Peysson2012}. However, the ray formulation is valid only when either the structure size is not the same order as the wavelength or the fluctuation level is not so large that the density gradient changes on the same length scale as the wave \cite{Fitzpatrick_book, Stix_book}. Therefore no one ray method includes the scattering effect of the full continuous $k$ spectrum (from small to large structures) of the turbulence nor the full spread in fluctuation amplitudes observed in a tokamak plasma edge. Though the assumptions above are always partially valid for some parts of the turbulence spectrum in the edge, it is certainly not the entire description and could lead to significant underestimates of the beam broadening.  

\vspace{5mm}

Full-wave Finite Difference Time Domain (FDTD) codes are not limited in the ways discussed above. Because this method directly solves Maxwell's equations it is referred to as full-wave and includes both the effects of multi scale turbulence and cross polarisation scattering within the formulation of the wave mechanics (see section \ref{Subsec:EMIT-3D}). Furthermore full-wave methods are able to use a more complete description of the turbulence, generated by specific fusion plasma turbulence codes. The output from gyrokinetic, gyrofluid or fluid models of a tokamak edge can be used as direct input (See section \ref{Subsec:Hermes}). Full wave simulations are generally computationally expensive compared to the ray methods described above and in the past have been used to look at specific problems which require a treatment using the full description of the wave mechanics. Most recently progress has been made in understanding the scattering effect of density perturbations on an incoming beam through a large range in turbulence parameter space \cite{Kohn2016, Kohn2017}. The results from these full-wave codes are often used to benchmark the ray codes with the inclusion of statistical turbulence. 

\vspace{5mm}

There are many predictions for ECCD broadening due to edge density fluctuations using beam tracing codes with a statistical turbulence approximation \cite{Decker2012, Tsironis2009, Bertelli2010, Sysoeva2015, Peysson2011, Peysson2012}. There are no similar studies using a full wave code along with a turbulence code to treat the edge. This makes benchmarking with our most complete theoretical description impossible. Furthermore there has only been one ray tracing study (with statistical turbulence included) which directly compares the modelled ECRH beam broadening with the experimentally measured heating profile, for a particular shot, on a systematic basis \cite{Decker2012}. As a consequence there have been no opportunities to test our theoretical models in multiple shots with substantially differing edge turbulence (once more there are no full wave equivalent). The purpose of this work is to address these issues. The paper is laid out in the following way: We begin in section \ref{Sec:Experiment} with the experimental measurements of ECRH broadening in DIII-D for the specific shots simulated by EMIT-3D. In section \ref{Sec:Simulation_Domain} the EMIT-3D code (section \ref{Subsec:EMIT-3D}) and turbulence model (section \ref{Subsec:Hermes}) are outlined along with the method for matching the simulation domain to experiment. Section \ref{Sec:Results} presents the analysis of simulation results and section \ref{Sec:Discus} provides a discussion of the work.

\section{Experimentally measured ECRH Deposition Broadening in DIII-D}
\label{Sec:Experiment}

\begin{figure}[H]
\centering
\includegraphics[width=0.4\linewidth]{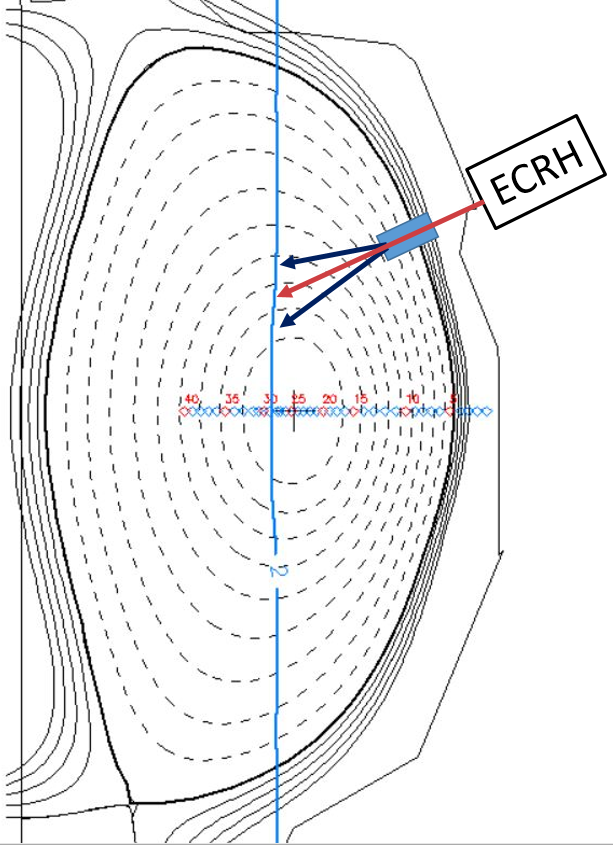}
\caption{An illustration of the ECRH launch angle and path of the rays. The blue rectangle approximately covers the full wave simulation domain and the arrows by the subsequent projection technique discussed in section \ref{Subsec:Extrapolation}. The blue vertical line is the targeted electron cyclotron resonance surface. The separatrix is denoted by the solid black line. The blue circles along the mid-plane illustrate the view of the diagnostics relative to the ECRH injection.}
\label{Fig:Simulation_domain_in_experiments}
\end{figure}

In DIII-D 110 GHz X-mode polarised beams are used to drive ECCD and ECRH at the location where the electron cyclotron frequency is 55 GHz (second harmonic absorption). Beams from up to 6 gyrotrons are launched from a set of steerable mirrors located at a poloidal angle of $60^{\circ}$. The beam waist radius at the launcher is between $6 - 7$ cm ($\approx 23 \lambda_0$) depending on the launcher, resulting in a Gaussian beam with a large waist to wavelength ratio which can be approximated as a plane wave solution with very little divergence. The beam path and deposition profile is calculated by the ray tracing code TORAY-GA on a shot by shot basis (see figure \ref{Fig:Simulation_domain_in_experiments}). The corresponding experimental work to this paper, described in detail in \cite{Brookman2017b}, has shown significant differences between the calculated deposition profile and the measured heating profile. It was found that the broadening of the heating profile cannot be explained by transport effects and instead is found to scale with the level of edge density fluctuation. This led to the conclusion that the beam broadening is caused by scattering of the beam in the tokamak edge region. The experimental broadening is defined as a ratio of the measured heating deposition width and that calculated by TORAY-GA and is shown in table \ref{Tab:Experimental_Measurements}.

\begin{table}[H]
\small
\begin{center}
    \begin{tabular}{ p{2.0 cm} | p{2.55 cm} | p{2.55 cm}|  p{1.75 cm} | p{2.55 cm} | p{2.55 cm} } \hline
        & Neg-Tri L- & Diverted H- & QH-mode & Limited L- & Diverted L- \\ 
           & mode $\#166191$ & mode $\#165146$ & $\#157131$ & mode $\#154532$ & mode $\#165078$ \\  \hline\hline
     
	Fluc. Level & 3 & 5 & 8 & 10 & 12 \\
	Broad. Fac. & $\times1.4\pm 0.2$ & $\times1.7\pm 0.2$ & $\times1.9\pm 0.4$ & $\times2.3\pm 0.2$ & $\times2.7\pm 0.3$ \\

    \end{tabular}
\end{center}
\caption{Plasma scenario and corresponding shot number with measurements of the fluctuation level made on the normalised poloidal flux surface $\psi_N = 0.95$ and the ECRH broadening factor. The fluctuation level is given as an RMS percentage fluctuation about the background density and is shown mathematically in equation \ref{eqn:RMS_fluc_level}. The H-mode shot is inter-ELM and so does not include scattering effects of ELMs.} 
\label{Tab:Experimental_Measurements}
\end{table}

\section{Theoretically modelled ECRH Deposition Broadening in DIII-D}
\label{Sec:Simulation_Domain}

\subsection{The EMIT-3D code}
\label{Subsec:EMIT-3D}

EMIT-3D utilises the finite difference time domain (FDTD) approach to solve Maxwell's equations along with a plasma response, shown below. The FDTD method in EMIT-3D makes use of the Yee cell for staggered spatial stepping and the leapfrog method for time stepping which together result in a second order accurate approach. The current density evolution equation couples the wave electric field to the plasma density through ($\omega_{pe}^2$ $= e^2n_e/m_e\epsilon_0$) and the current density to the background plasma magnetic field, $B_0$, via ($\omega_{ce}$ $= eB_0/m_e$). Once the equations are linearised all field vectors \textbf{E}, \textbf{B} and \textbf{J} become the perturbing fields associated with the wave since the background plasma fields are taken to be in equilibrium on the time-scales of the wave and therefore drop out of the equations. The background magnetic field is assumed to be spatially slowly varying in relation to the wavelength and so the background current density does not enter the equations. In this version the \textbf{B} and \textbf{J}-fields are updated one half time step after the \textbf{E}-field.

\begin{equation}
\label{eqn:Maxwell}
\begin{aligned}
\frac{\partial \textbf{E}}{\partial t} &= \frac{1}{\mu_0 \epsilon_0} \nabla \times \textbf{B} - \frac{1}{\epsilon_0} \textbf{J} \\
\frac{\partial \textbf{B}}{\partial t} &= -\nabla \times \textbf{E} \\
\frac{\partial \textbf{J}}{\partial t} &= \epsilon_{0}\omega_{pe}^{2}\textbf{E} - \omega_{ce}\textbf{J} \times \hat{\textbf{b}}_{0}
\end{aligned}
\end{equation}

The boundaries use a damping layer to reduce the wave amplitude to zero thus simulating an infinite box. Within the boundary layer the wave electric field is multiplied by a parabolic function $D(r)$ of the form $D(r) = 1 + \frac{13}{T}((r - d_{bound})/d_{bound})^3$, where T is the wave period, $d_{bound}$ is the boundary thickness and $r \leq d_{bound}$. The boundary layer is three vacuum wavelengths thick after which the wave is reflected resulting in a total propagation distance of six vacuum wavelengths over which the wave is damped. The antenna is launched as a 2D Gaussian beam which excites a single directional component of the wave electric field and can be phased across the array so that the resulting beam propagates at any angle chosen. By carefully selecting the orientation of the magnetic field, the excitation component and the beam angle, one can excite pure single polarisation modes or any specific mixture of polarisations.

EMIT-3D is fully parallelised in three dimensions. The global 3D cuboid shaped grid is split into a number of smaller cuboids equal to the number of processes. Each cuboid is created on its allocated process and therefore only holds data for its portion of the grid. The processes communicate with adjacent processes between each leapfrog update time-step. In this way the \textbf{E}-field data is communicated first after its update followed by the \textbf{B} and \textbf{J}-fields one half time-step later after their update. The communication is achieved via single layer ghost cells which surround the cuboid and do not themselves update. The scalability for this method of parallelisation results in a near doubling of the speed when doubling the number of cores. Notable loss in this scalability trend is only encountered when a large fraction of the split cuboid is taken up by ghost cells ($>$ 10\%).

In an earlier form EMIT-3D has been used to study the effect of single blobs on propagating electromagnetic waves \cite{Williams2014}. More recently EMIT-3D has been used in conjunction with another full wave code to investigate the scattering effect of turbulence over a large range in turbulence parameter space \cite{Kohn2016, Kohn2017}.

\subsection{Ensuring the simulated beam and plasma match the experimental conditions}

\subsubsection{The simulation domain and beam initialisation}
\label{Subsec:Domain}

\begin{figure}[H]
\centering
\begin{subfigure}{.5\textwidth}
	\includegraphics[width=\linewidth]{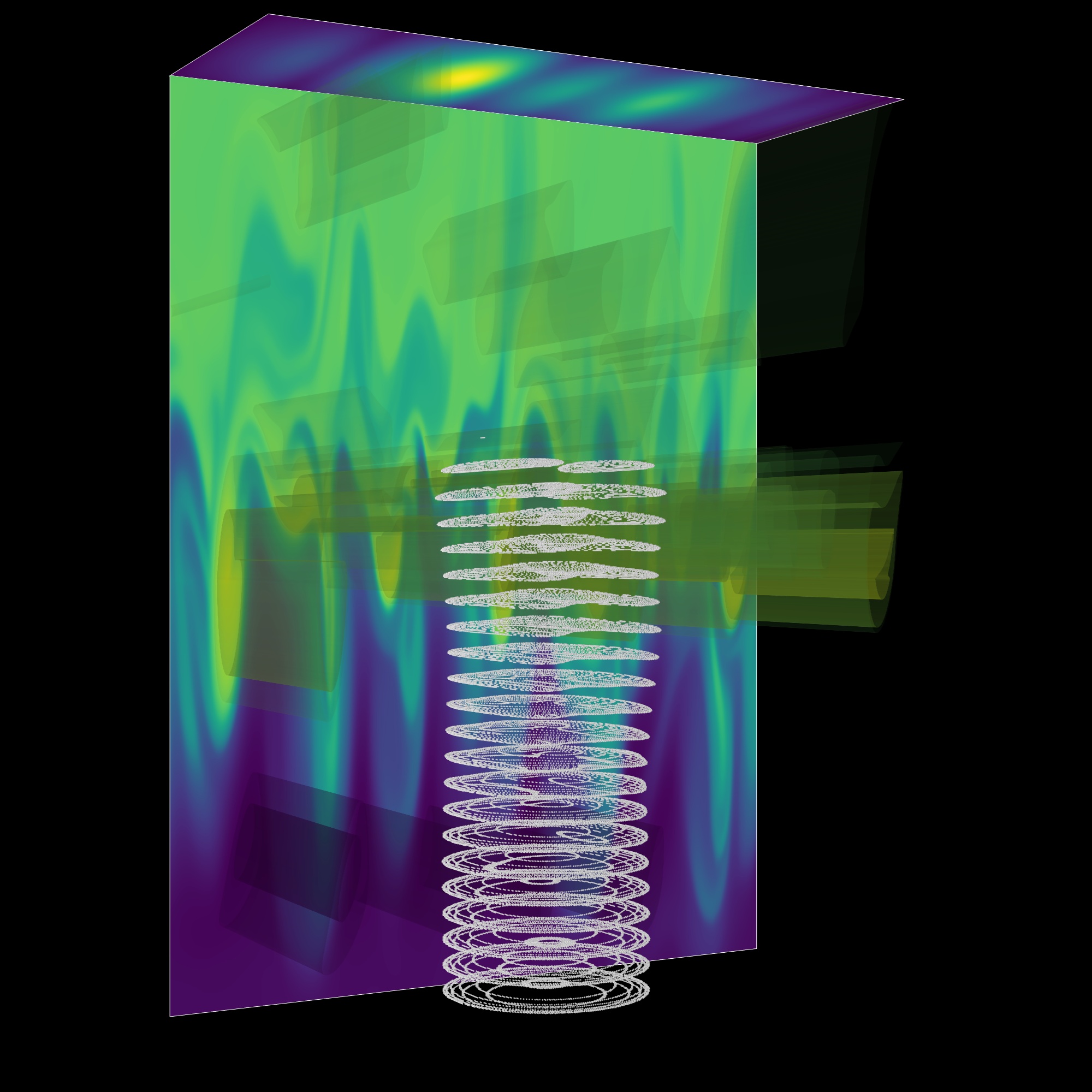}
\end{subfigure}%
\begin{subfigure}{.5\textwidth}
	\includegraphics[width=\linewidth]{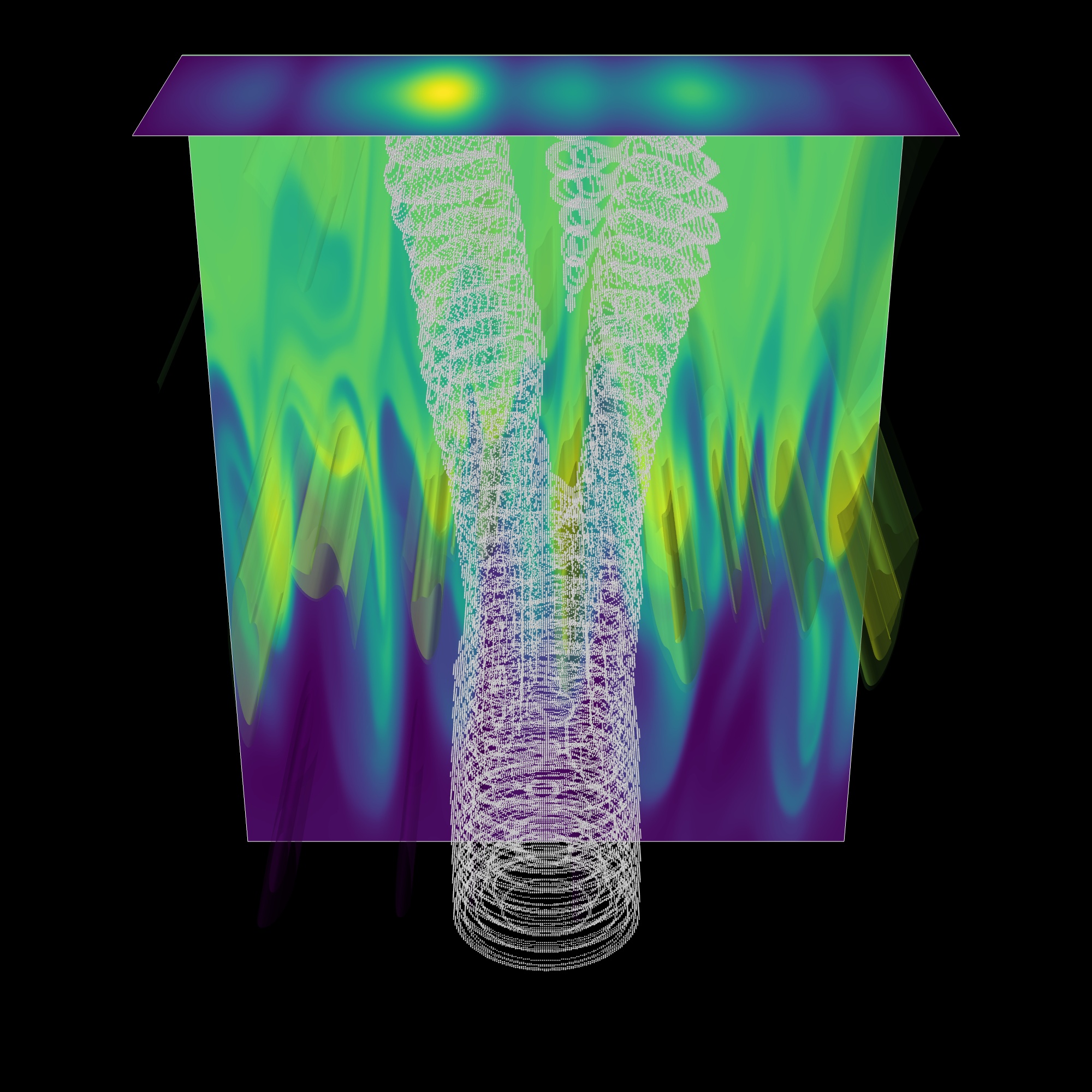}
\end{subfigure}
\caption{The output from EMIT-3D for one particular turbulence snapshot in in the ensemble for L-mode. Left: The 2D turbulence profile generated by BOUT++ is shown to be extended along the magnetic field lines into filamentary structures and a 3D wave propagates from low density to high density (blue to green). The white contours are plotted at the beam width where the amplitude falls to $1/e$ of the maximum. Right: A later time in the same simulation. The image at the backplane (top) is the RMS electric field calculated over two wave periods.}
\label{Fig:Simulation_Domain}
\end{figure}

Of the cases shown in table \ref{Tab:Experimental_Measurements} three cases with distinct edge plasma characteristics, L-mode, H-mode and negative triangularity L-mode, are considered to directly compare with experiment. As discussed in section \ref{Sec:Intro} a full wave treatment is only necessary when the inhomogeneity scale length is comparable to the wavelength. This occurs in the steep density gradient region of the plasma edge. Because full wave simulation is computationally expensive, the simulation domain is chosen to span this region about the separatrix shown in figure \ref{Fig:Simulation_domain_in_experiments}. The method for propagating to the absorption surface from the end of the simulation domain is discussed in section \ref{Subsec:Extrapolation}. To reduce computational expense the domain size was constrained to be as small as possible without affecting the wave mechanics. To this end a resolution of 20 Yee cell grid points per vacuum wavelength is chosen. The optimum width of the simulation domain was found to be 4 times the beam diameter in the direction perpendicular to the filamentary structures in $\hat{x}$ (where scattering is expected) and 1.8 in the direction parallel to filamentary structures in $\hat{y}$ (figure \ref{Fig:Simulation_Domain}). The experimental beam waist is 47$\lambda_0$ which is not practical to simulate here. Instead a beam is launched which ensures a flat plane wave-front. It was found in previous work \cite{Kohn2016} that the dependence of scattering power of turbulence on a Gaussian beam asymptotes to a constant value beyond a beam waist 6$\lambda_0$. Therefore 6.8$\lambda_0$ was chosen to represent the experimental beam. An X-mode Gaussian beam is launched by exciting only the component of the wave electric-field that is perpendicular to both the background magnetic field and the wave vector, $\hat{\textbf{k}}$, within a homogeneous background plasma. The background plasma is then smoothly transitioned to the turbulence profile by use of a hyperbolic tangent function. The turbulence layer was $38\lambda_0$ in the propagation direction $\hat{z}$.

\subsubsection{Generation of the turbulence}
\label{Subsec:Hermes}

The turbulence used for simulating microwave interaction was generated using the BOUT++ framework \cite{Dudson2009}. The Hermes model \cite{Dudson2017} was simplified to include only electrostatic effects in a quasi-3D geometry. In the perpendicular plane the fluid equations were modelled explicitly, while separate closures were used for the parallel direction to simulate the behaviour inside and outside the separatrix. The Hasegawa-Wakatani \cite{Hasegawa1983} closure describes the closed field line region, which models nearly adiabatic electrons through a parallel current and resistivity. Outside the separatrix, a sheath model provides a sink for plasma density and energy \cite{Strangeby_book}.

\subsubsection{Using experimental diagnostics to constrain the simulated turbulence characteristics}
\label{Subsec:Turbulence_analysis}

We begin with the diverted L-mode shot $\# 165078$. The Thomson Scattering diagnostic \cite{Carlstrom1992} provides the time averaged electron density (henceforth referred to as density) and electron temperature profile whilst EFIT provides the magnetic field. These serve as input to the Hermes model and drive the plasma turbulence. As the frequency of the wave is much larger than the eddy turnover frequency, the turbulence is considered as stationary. Therefore, as the turbulence evolves in time, 2D density profile snapshots are selected at sufficiently spaced time intervals to provide a large uncorrelated ensemble in order to capture the time averaged effect of turbulence on the beam. Away from the divertor the parallel density gradients are negligible due to fast parallel transport. This requires that the 2D profile is then extended into 3D along the magnetic field lines reproducing the distinctive filamentary structures. The ensemble characteristics are then cross checked with experiment to ensure that the spatial distribution of the root mean square (RMS) fluctuation level $\xi$, time averaged density $n_{e,0}(r)$, poloidal and radial correlation lengths are matched. The RMS fluctuation level is obtained by beam emission spectroscopy (BES) \cite{McKee1999}. It is defined in equation \ref{eqn:RMS_fluc_level} as the RMS fluctuation amplitude normalised to the time averaged density. 

\begin{equation}
\label{eqn:RMS_fluc_level}
\xi = \sqrt{\frac{1}{N}\sum_{t=1}^{N}\left(\frac{n_e(r,t) - n_{e,0}(r)}{n_{e,0}(r)}\right)^2} 
\end{equation}

\begin{figure}[H]
\includegraphics[width=\linewidth, height = 100mm]{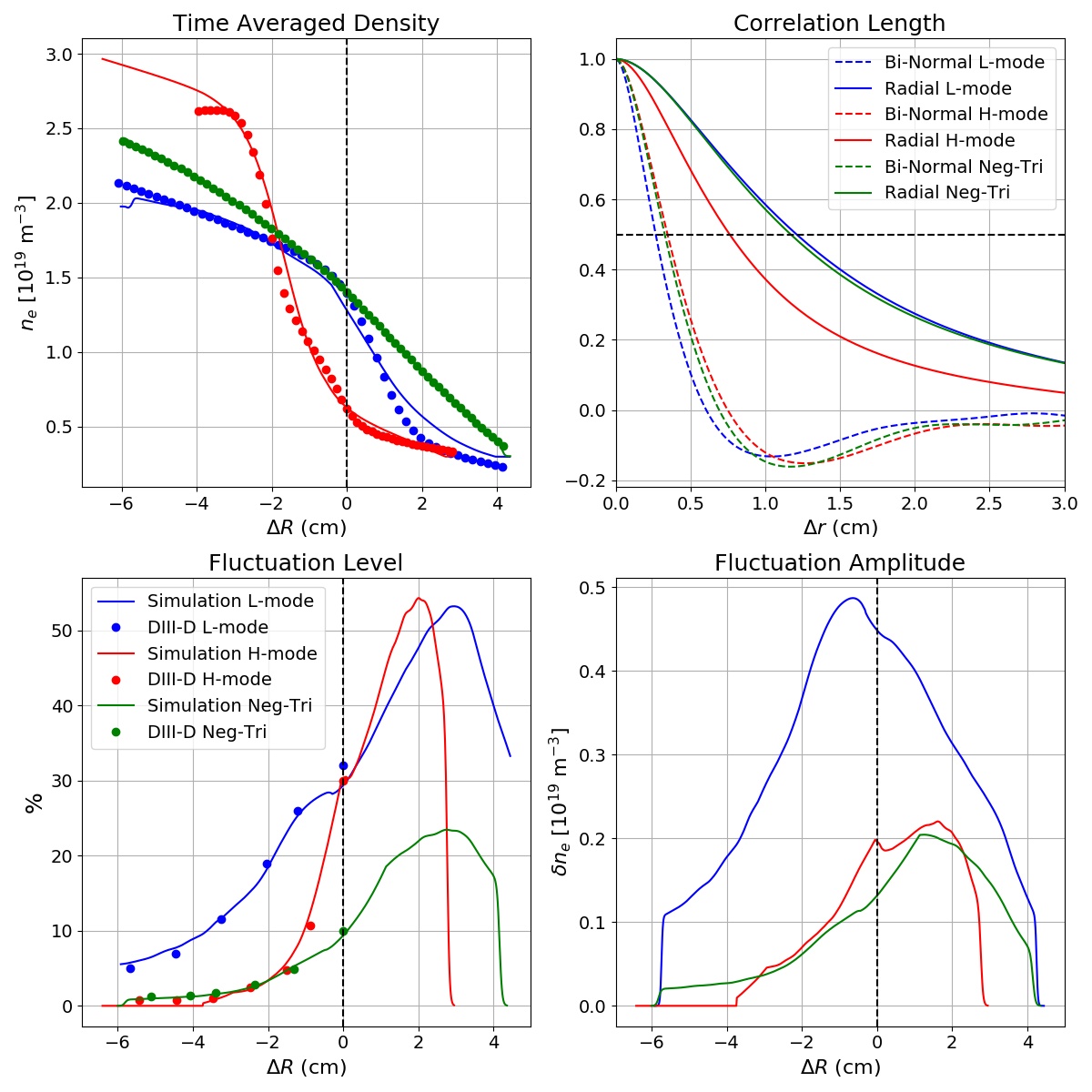}
\caption{The properties of the generated turbulence ensemble along with the corresponding DIII-D measurements for each scenario considered. The dots represent experimental measurements whilst the solid lines are from the simulated turbulence. Only the left two plots contain the experimental data. Each scenario is colour coded as shown in the legend of the bottom left plot. $\Delta R$ is with respect to the outboard mid-plane with 0 denoting the separatrix.}
\label{Fig:Turbulence_analysis_profiles}
\end{figure}

N is the ensemble size and $n_e(r,t)$ denotes the density profile in $r$ for an instantaneous time $t$. The RMS fluctuation amplitude is defined as $\tilde{n}_e(r) = \sqrt{\sum\left(n_e(r,t)-n_{e,0}(r)\right)^2/N}$. The correlation length, $L_c$, is defined as the average distance over which the normalised autocorrelation falls off to 0.5 which corresponds to the eddy radius when the density fluctuation amplitude falls to 1/e. This is clearly an averaged quantity and a range of structure sizes of differing power are present in the profile. The physics governing the correlation lengths is vastly different over the radial extent of the simulation domain spanning the separatrix. In the scrape off layer (SOL) blobs dominate the structure therefore the radial correlation length is dominated by larger length scales \cite{Devynck2005, Grulke2006} and across the separatrix the simulations show that radial length scales are enlarged due to the radial direction of advection. By contrast, inside the separatrix the binormal correlation length has been measured to be of the same size \cite{Jakubowski2002} and even larger \cite{McKee2001}. It is difficult to exactly define the ratio between the radial and poloidal length scales on average in experiment due to the sensitivity of different diagnostics to specific values of $k$ and also due to the spatial and temporal resolution. What can be stated though is a range in which the measured correlation lengths lie for DIII-D \cite{McKee2007, Shafer2012, Zweben2007}. This is used as the constraint for the simulations. Furthermore as the correlation length only describes an average quantity it is necessary to check the $k$ power spectrum of the simulated turbulence. The spectrum follows known power laws in specific regions corresponding to different underlying turbulence mechanisms, a comprehensive review of which can be found in Ref. \cite{Hennequin2006}. Figure \ref{Fig:Simulation_Domain} illustrates the simulation domain with the turbulence generated from the Hermes model and the EMIT-3D output. Figure \ref{Fig:Turbulence_analysis_profiles} shows the simulated and experimentally measured parameters described above for each of the cases modelled. The BES data for the RMS fluctuation level is only present inside the separatrix. Outside the separatrix it is understood that the fluctuation level continues to increase up to levels of 60\% and in some cases 100\%. A comprehensive review of edge turbulence measurements in toroidal fusion devices \cite{Zweben2007} demonstrates this increase in the SOL. The fluctuation levels and correlation lengths are also consistent with previous DIII-D measurements \cite{Hennequin2006,McKee2007,Shafer2012,Zweben2007}.

\vspace{5mm}

The turbulence mechanism for the negative triangularity L-mode operating scenario is the same as that for normal L-mode configuration though significantly suppressed \cite{Marinoni2009}. Therefore the same profiles generated for the normal L-mode scenario were scaled to fit the negative triangularity experimental data. The time averaged density is changed by dividing through each snapshot with the L-mode time averaged density and multiplying through by the fit to the negative triangularity data. In a similar way once the time averaged density has been removed, leaving only the fluctuations, the RMS fluctuation level can be manipulated using a fit to the BES data for the required shot. The correlation length is altered through interpolation of the grid to span a larger or shorter distance relative to the wave as required. Ideally for H-mode a multi scale gyrokinetics simulation of DIII-D would be used. However this was not available due to the huge computational expense of such a simulation at a grid resolution small enough for the wave mechanics to fully interact with the fine structure. Therefore the turbulence profiles generated by Hermes were scaled as described above to be consistent with the measured turbulence parameters of the H-mode shot. The modelled H-mode turbulence does not include ELMs because the experimental measurement is inter-ELM. The radial correlation length in H-mode is expected to be reduced in comparison to L-mode due to the $\textbf{E} \times \textbf{B}$ shear flow whilst increasing in the binormal direction \cite{Hennequin2006, Conway2005, Schirmer2007}. For negative triangularity the correlation lengths are generally unchanged from L-mode according to the dimensionless $\rho^{*}=\rho_i/a$ scaling law previously shown in \cite{McKee2001, Hennequin2006}. The experimental and simulated parameters for each case are shown in figure \ref{Fig:Turbulence_analysis_profiles}.

It is noted that an unavoidable challenge between experiment and simulation is encountered. The ECRH beam launcher is located at a poloidal angle of $60^{\circ}$ whereas the diagnostic data used to match the simulation to experiment is located on the mid-plane. Recent global simulations of DIII-D \cite{Dudson2017, Cohen2013} have demonstrated that the turbulence amplitude changes with poloidal angle. The change is either positive or negative depending on the turbulence model used and the operating regime considered. Therefore it is not understood whether the change occurs in the same way for the total $k$ spectrum of the turbulence or part of it. What can be concluded though is because the change in angle from the mid-plane is small (see figure \ref{Fig:Simulation_domain_in_experiments}) one should expect only a slight difference between the broadening estimated here and the measured deposition profile broadening.

\section{Results}
\label{Sec:Results}

\subsection{The ensemble}
\label{Ensemble}
For each modelled scenario a reference case is needed where the beam propagates through the time averaged density profile without turbulence. When turbulence is added a large ensemble of simulations per scenario is required. The root mean square of the wave electric field (equation \ref{eqn:RMS_Electric_Field}) is calculated for each snapshot which is used to define a scatter parameter, $\alpha_s$ (equation \ref{eqn:Scatter_Parameter}), which can be physically interpreted as a measure of the percentage change in the distribution of the beam energy across the two dimensional grid at the back-plane. The back-plane is shown in figure \ref{Fig:Simulation_Domain}. 

\begin{equation}
\label{eqn:RMS_Electric_Field}
E_{RMS} = \sqrt{\frac{1}{2T}\sum_{t=1}^{2T}\left(E_{x,t}^2 + E_{y,t}^2 + E_{z,t}^2\right)}
\end{equation}

\begin{equation}
\label{eqn:Scatter_Parameter}
\alpha_{s} = \frac{ \sum_{i,j} \left( \widetilde{E^2}_{RMS}(i,j) - E_{RMS}^2(i,j) \right) }{\sum_{i,j} E_{RMS}^2(i,j)}
\end{equation}

\begin{equation}
\label{eqn:Standard_deviation}
\sigma_{ensemble} = \sqrt{\frac{1}{N}\sum^{N}_{s=1}(\alpha_{s} - mean(\alpha_{s}))^2}
\end{equation}

\begin{figure}[H]
\centering
\includegraphics[width=100mm, height = 70mm]{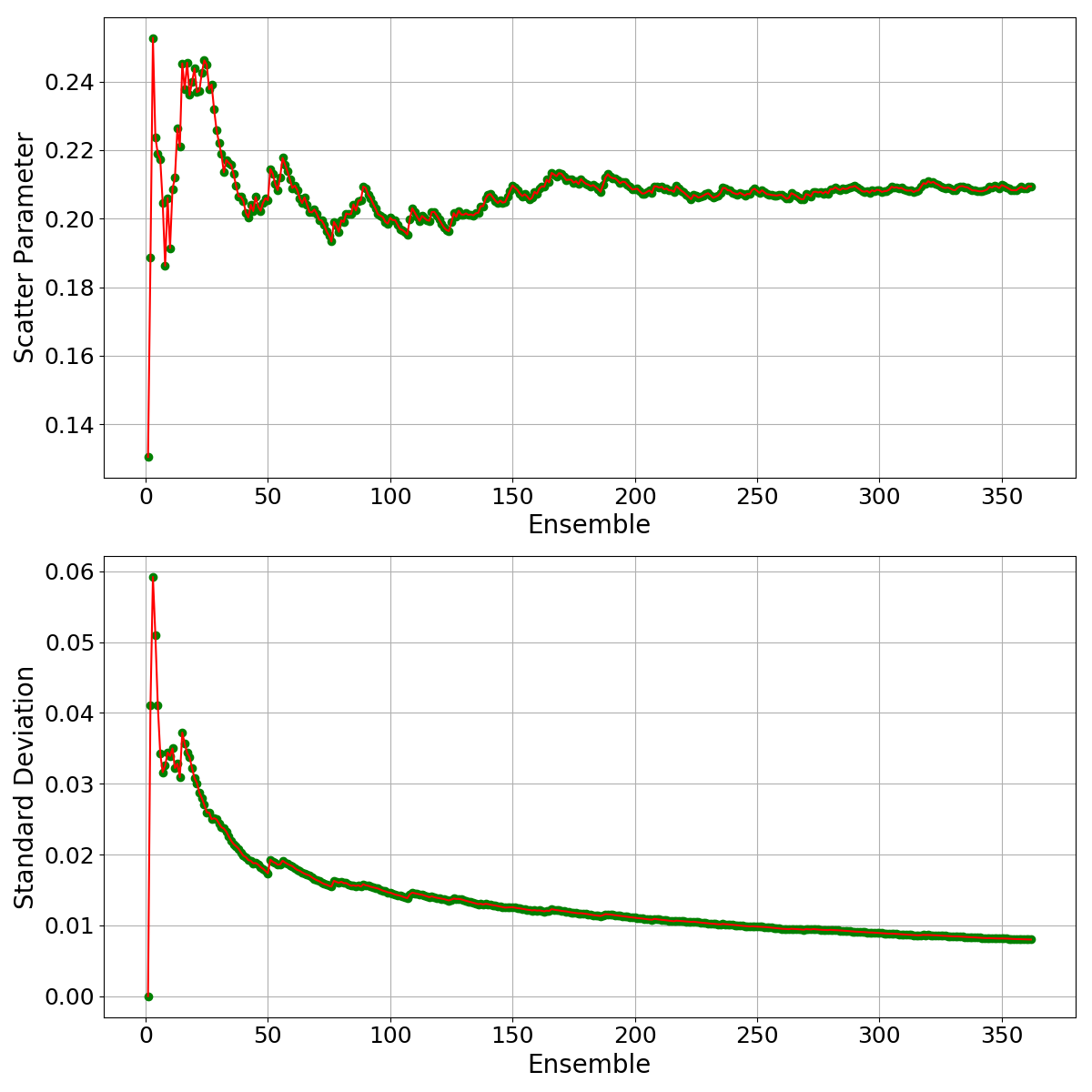}
\caption{Convergence of the scatter parameter $\alpha$ and the standard deviation $\sigma$ with increasing ensemble size for L-mode.}
\label{Fig:Lmode_convergence}
\end{figure}

The necessary ensemble size may be determined by looking for the convergence of both the mean of $\alpha_s$ over the ensemble (denoted $\alpha$) and also its standard deviation $\sigma$ (figure \ref{Fig:Lmode_convergence}). For L-mode, convergence was found after approximately 250 simulations though 360 were used in the total ensemble. For H-mode and negative triangularity, convergence was found much sooner because the turbulence was weaker and therefore did not perturb the beam to the same extent. This allowed a smaller ensemble size of 150 to be used.

\begin{figure}[H]
\centering
\begin{subfigure}{.75\textwidth}
	\includegraphics[width=\linewidth]{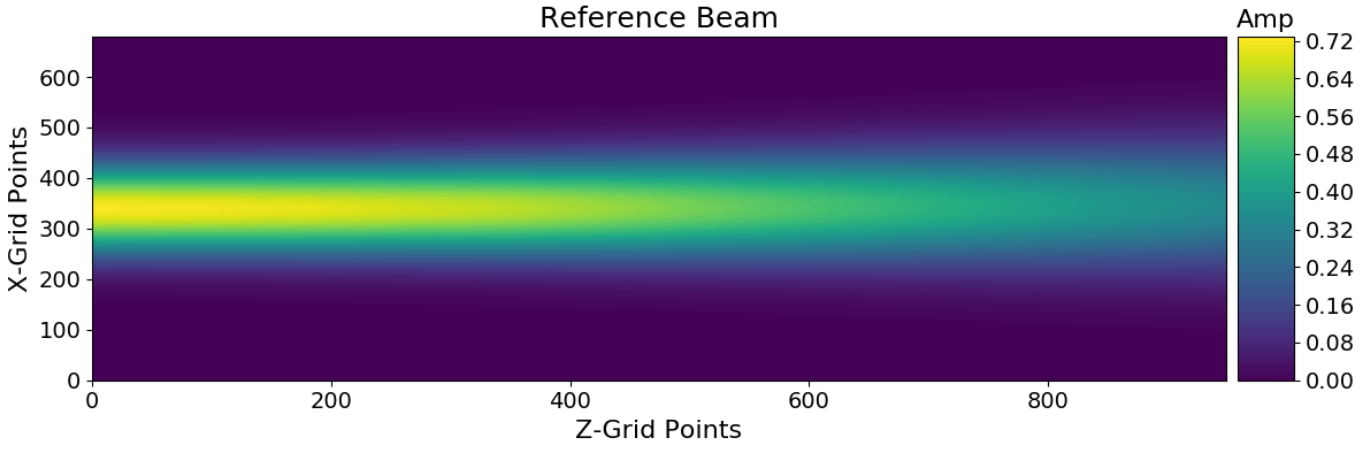}
\end{subfigure}%
\begin{subfigure}{.25\textwidth}
	\includegraphics[width=\linewidth, height=40mm]{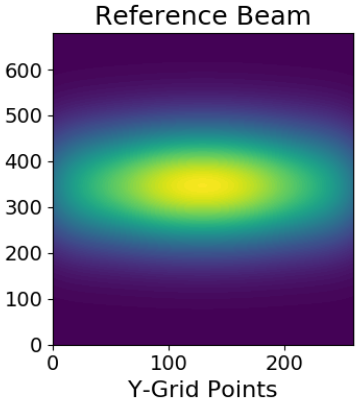}
\end{subfigure}

\begin{subfigure}{.75\textwidth}
	\includegraphics[width=\linewidth]{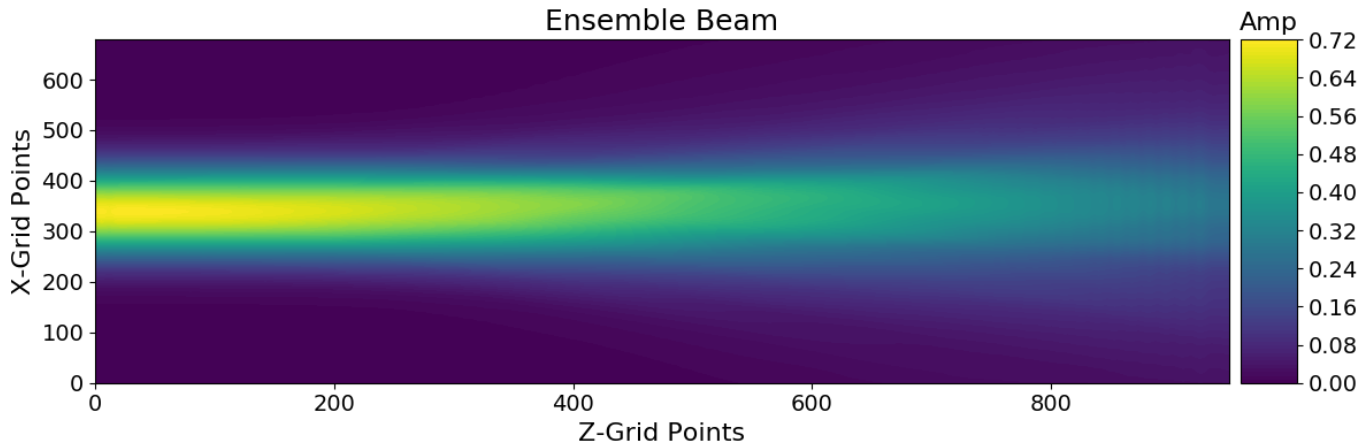}
\end{subfigure}%
\begin{subfigure}{.25\textwidth}
	\includegraphics[width=\linewidth, height=40mm]{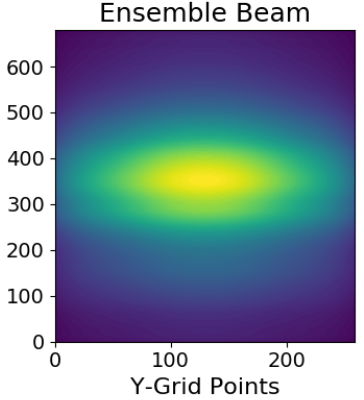}
\end{subfigure}
\caption{Output from the L-mode simulations. Left: The RMS electric field of the ensemble beam and the reference beam along the propagation direction. Right: A cross section through the beam at the back-plane.}
\label{Fig:Lmode_Reference_and_Ensemble_Beam}
\end{figure}

Once these parameters are observed to sufficiently converge one may take the mean of the RMS wave electric field along the beam propagation to analyse the time averaged effect of the beam over the ensemble. The result is a Gaussian beam with increased width and divergence in relation to the reference case due to scattering from turbulence (figure  \ref{Fig:Lmode_Reference_and_Ensemble_Beam}). However it is noted that though the main body of a time averaged scattered beam fits very well to a Gaussian after, side lobes develop which are not captured by the Gaussian function; these will be discussed later.

\subsection{Projection to the absorption surface}
\label{Subsec:Extrapolation}
The absorption surface of the ECCD power is much further into the plasma than the edge of the simulation domain. The exact distance is case specific. Due to the difference in divergence, the ratio of the beam widths will not remain constant from the simulation domain back-plane to the absorption surface. Therefore a method is required to calculate the ratio at the absorption surface given the divergence of both beams at the end of the simulation domain. The envelope of a Gaussian beam in homogeneous space after the Rayleigh range ($R_R = \pi w(0)^2 / \lambda_0$) will asymptote to a straight line drawn to infinity from the centre of the beam at the waist, at an angle corresponding to the divergence of the beam. This quality of a Gaussian beam allows one to fit straight lines to the envelope of the beam after the Rayleigh range to extrapolate the beam to greater distances assuming the subsequent propagation is through homogeneous space. This can be done for both the reference beam and the ensemble scattered beam as a first order approximation. This also assumes that the ensemble beam continues to be well described by a Gaussian whose divergence is defined by the ensemble scattering in the simulation domain. 

\vspace{5mm}

The limitation of the method is that no further refractive effects from the changing density gradient and magnetic field vector will be included. A further broadening of the beam is expected due to the orientation of the beam $k$ vectors distributed over the wave-front to the density gradient vector and magnetic field vector. This effect is expected to be small due to the length scale of the gradients in comparison with the wavelength and will affect both the reference and ensemble beam in the same way. However the effect will be larger for the ensemble beam due to its larger divergence and although it will result in an underestimation of the broadening at the absorption point it is not thought to contribute by more than 10\% as a conservative estimate.

\subsection{Beam broadening in each scenario}
\label{Subsec:Broadening}
\begin{figure}[H]
\centering
\begin{subfigure}{.25\textwidth}
	\includegraphics[width=\linewidth, height=40mm]{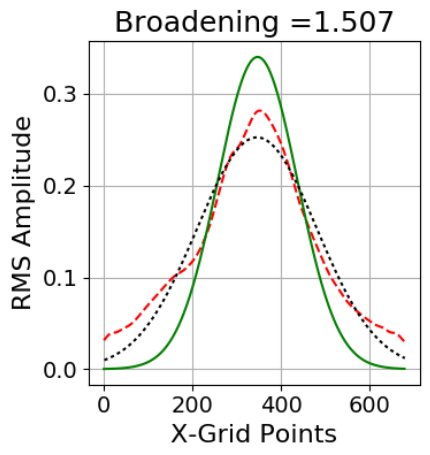}
\end{subfigure}%
\begin{subfigure}{.75\textwidth}
	\includegraphics[width=\linewidth]{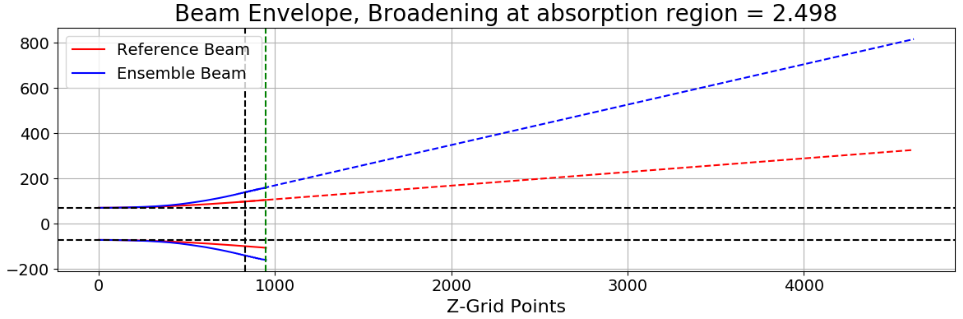}
\end{subfigure}

\begin{subfigure}{.25\textwidth}
	\includegraphics[width=\linewidth, height=40mm]{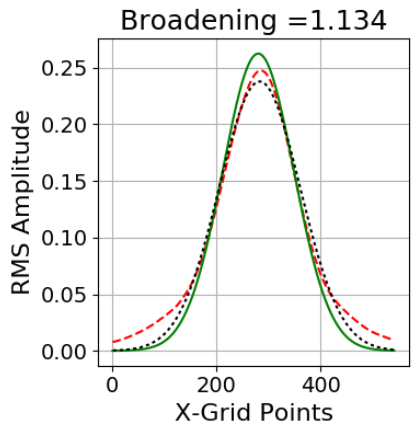}
\end{subfigure}%
\begin{subfigure}{.75\textwidth}
	\includegraphics[width=\linewidth]{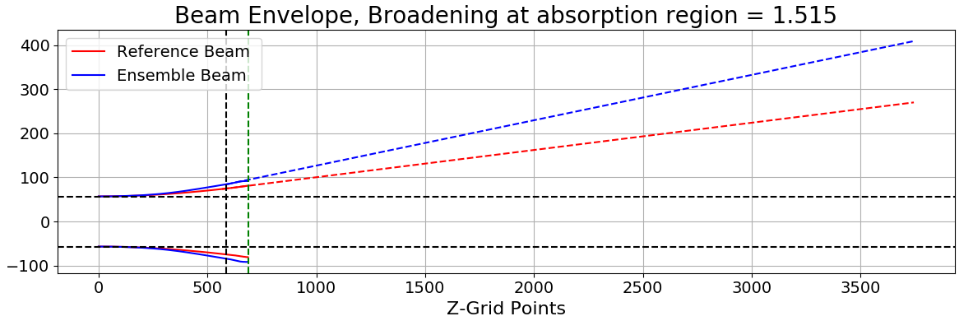}
\end{subfigure}

\begin{subfigure}{.25\textwidth}
	\includegraphics[width=\linewidth, height=40mm]{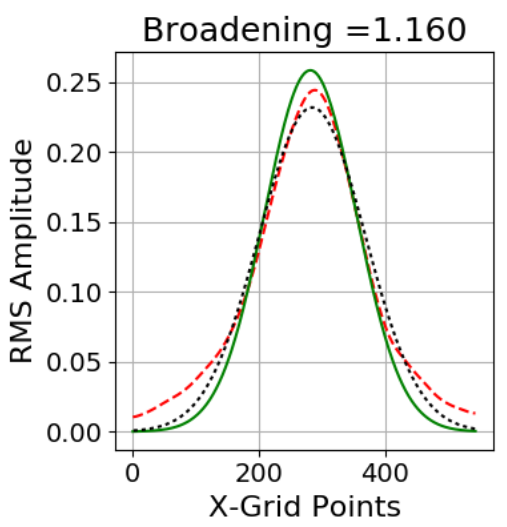}
\end{subfigure}%
\begin{subfigure}{.75\textwidth}
	\includegraphics[width=\linewidth]{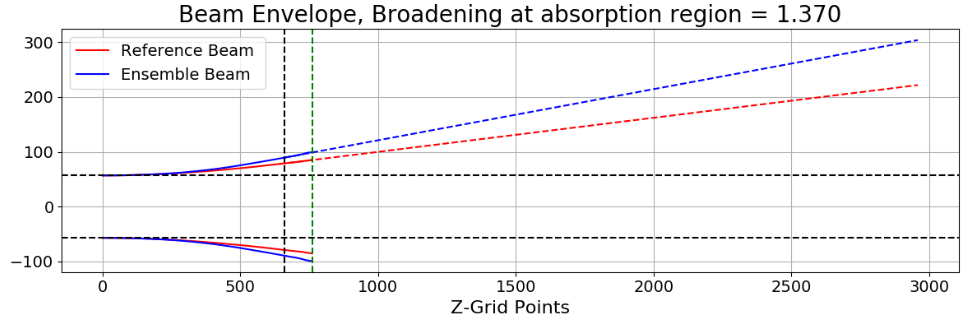}
\end{subfigure}
\caption{The top two plots are the L-mode results, the middle two are H-mode and the bottom two are negative triangularity. Left: A line plot of the back-plane beam cross section taken in the scattering direction. The backplane location and the scattering direction is shown as the green dashed line on the right hand plot. Shown is the RMS electric field for both the reference beam (green), the ensemble beam (red dashed) and the Gaussian fit (black dashed) to the ensemble beam. Right: Extrapolation of the beam envelope to the absorption region from the end of the simulation domain (green dashed line). The extrapolated line is fit between the black and green dashed lines at the end of the simulation domain.}
\label{Fig:Broadening_extrapolation}
\end{figure}

Figure \ref{Fig:Broadening_extrapolation} shows the results of the three modelled scenarios and table \ref{Tab:Experimental_and_Smulation_results} compares these with the experimental measurements. For each of the scenarios in figure \ref{Fig:Broadening_extrapolation} one can see the full scale of the beam paths. The solid blue and red lines show the beam envelope of the ensemble and reference beams respectively inside the simulation domain which ends at the vertical green dashed line. The dashed blue and red lines then show the extrapolation of the beams as discussed in section \ref{Subsec:Extrapolation}. The lines are fitted according to the divergence of the two beams at the end of the simulation domain between the black and green dashed lines. The path length to the absorption surface from the end of the simulation domain is different in each case. In L-mode this was 40.1 cm ($\approx 147\lambda_0$), in H-mode it was 41.7 cm ($\approx 153\lambda_0$) and in negative triangularity it was 30.0 cm ($\approx 110\lambda_0$). It must be noted for comparing each graph directly, that for L-mode, a resolution of 25 grid points per wavelength was used and this was decreased to 20 in all other simulations.

In each case it is noted that side lobes develop which is an inherent property of scattering from turbulence and not dependent on the ensemble size. The side lobes are a direct result of the scattering from turbulent fluctuations having correlation lengths similar to the vacuum wavelength of the injected microwave and/or having a significant amplitude. This effect would not be expected in turbulence that is smoothly varying relative to the wavelength. It has been shown in \cite{Kohn2017} that the magnitude of scattering due to turbulence follows a log-normal distribution. The consequence is the occurrence of random large scattering events which significantly perturb the beam. The effect on the time-averaged beam is the appearance of side lobes which deviate from the Gaussian normal distribution. The ramification for the experiment is that more power may fall outside the width calculated here. 

\vspace{5mm}

The broadening effect of the edge plasma (i.e that calculated in the simulation domain only) is slightly more in negative triangularity than it is in H-mode despite the fluctuation level of the turbulence being larger in H-mode. This is due to the higher background density in negative triangularity combined with the much shallower gradient. This means that the actual fluctuation amplitude is the same as in H-mode and actually remains higher for a longer distance (see figure \ref{Fig:Turbulence_analysis_profiles}). The result is therefore not surprising and one can see that the reason the beam broadening is smaller at the absorption region is due to the much smaller propagation distance. It follows that the scaling of the expected beam broadening cannot simply be attributed to a measure of the fluctuation level at the edge but a combination of the background density and the fluctuation level. Moreover, the shape of the density profile and the fluctuation level profile play a major part in the magnitude of the beam broadening. This will be discussed further in the following section.

\begin{table}[H]
\small
\begin{center}
    \begin{tabular}{ p{3.2 cm} | p{3.0 cm} | p{3.0 cm}| p{3.0 cm} } \hline
        & Neg-Tri L-mode & Diverted H-mode & Diverted L-mode \\ 
           & $\#166191$ & $\#165146$ & $\#165078$ \\  \hline\hline
     
	Exp. Broad. Fac. & $\times1.4 \pm 0.2$ & $\times1.7 \pm 0.2$ & $\times2.7 \pm 0.3$ \\
	Sim. Broad. Fac. & $\times1.370 \pm 0.074$ & $\times1.515 \pm 0.102$ & $\times2.498 \pm 0.298$ \\
	
    \end{tabular}
\end{center}
\caption{The experimental measurements of the ECRH deposition profile broadening and the calculated beam broadening from the modelled ECRH beam in the same shot.} 
\label{Tab:Experimental_and_Smulation_results}
\end{table}

The errors on the calculation of the beam broadening from the simulation side are broken down as follows. The standard deviation is seen to converge between $\pm6\%$ to $\pm7\%$ depending on simulation scenario therefore contributing the same amount to the uncertainty on the beam broadening. A further contribution to the uncertainty as discussed in section \ref{Subsec:Extrapolation} comes from the method to extrapolate the beam envelope to the absorption region. This contributes $+10\%$ from the difference in refraction that will be encountered by the two beams through the core. A further $\pm7\%$ comes from the uncertainty in the fit of the line to the limited number of wavelengths at the end of the simulation domain and is calculated as the spread in broadening encountered. The total error is a combination of two indeterminate errors for which the sign is unknown and one determinate error with positive sign. When added together the indeterminate errors sum in quadrature and the determinate error sums conventionally. Consequently, the total error on the calculated beam broadening is $\sigma = \sqrt{7^2 + 7^2} + 10 = 19.9\%$. The uncertainty due to the mismatch in diagnostic position and ECRH launch position is unknown and therefore noted but omitted.

\vspace{5mm}

One can see from table 2 that the simulated beam broadening agrees with the experimental results very well and within the errors on the calculations. However there is a consistent under-prediction of the broadening. There is a slight discrepancy between the turbulence encountered by the experimental beam and the simulated beam due to the location of the diagnostics on the mid-plane and the injection angle at $60^{\circ}$ as discussed in section \ref{Subsec:Hermes}. However it is uncertain whether this would lead to an under or over-prediction of the broadening. Most likely the consistent nature of the under-prediction could be partially attributed to the fast electron transport which is not decoupled from the experimental measurement and is contained within the errors. This is consistent with the constraint on the fast electron transport to contribute no more than 10\% to the deposition broadening \cite{Casson2015}. Furthermore, as discussed in section \ref{Subsec:Extrapolation}, the method for projecting to the absorption surface is predicted to underestimate the broadening by no more than 10\% as a conservative estimate, which is contained in the simulation errors. These two effects combine to explain the consistent under-prediction. L-mode and negative triangularity simulations both agree with experiment to the order of 10\%, whereas H-mode is of the order 25\%. The reason for this extra 15\% may be attributed to the resolution of the diagnostics which affects the H-mode simulation the most due to the large gradients in the edge. This is discussed in the following section.

\subsection{Sensitivity of the modelled broadening to resolution of diagnostics}

\begin{figure}[H]
\centering
\begin{subfigure}{.33\textwidth}
	\includegraphics[width=\linewidth]{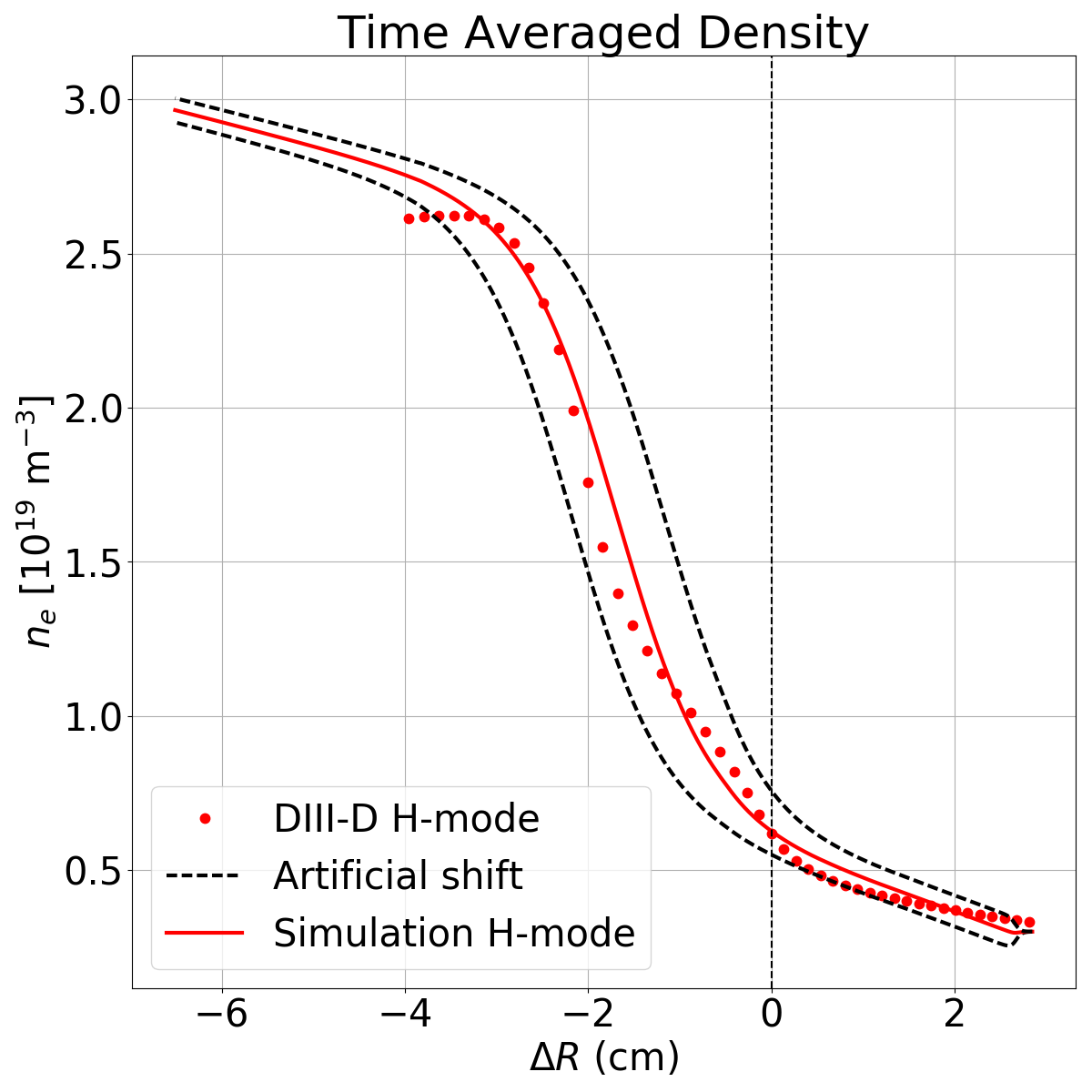}
\end{subfigure}%
\begin{subfigure}{.33\textwidth}
	\includegraphics[width=\linewidth]{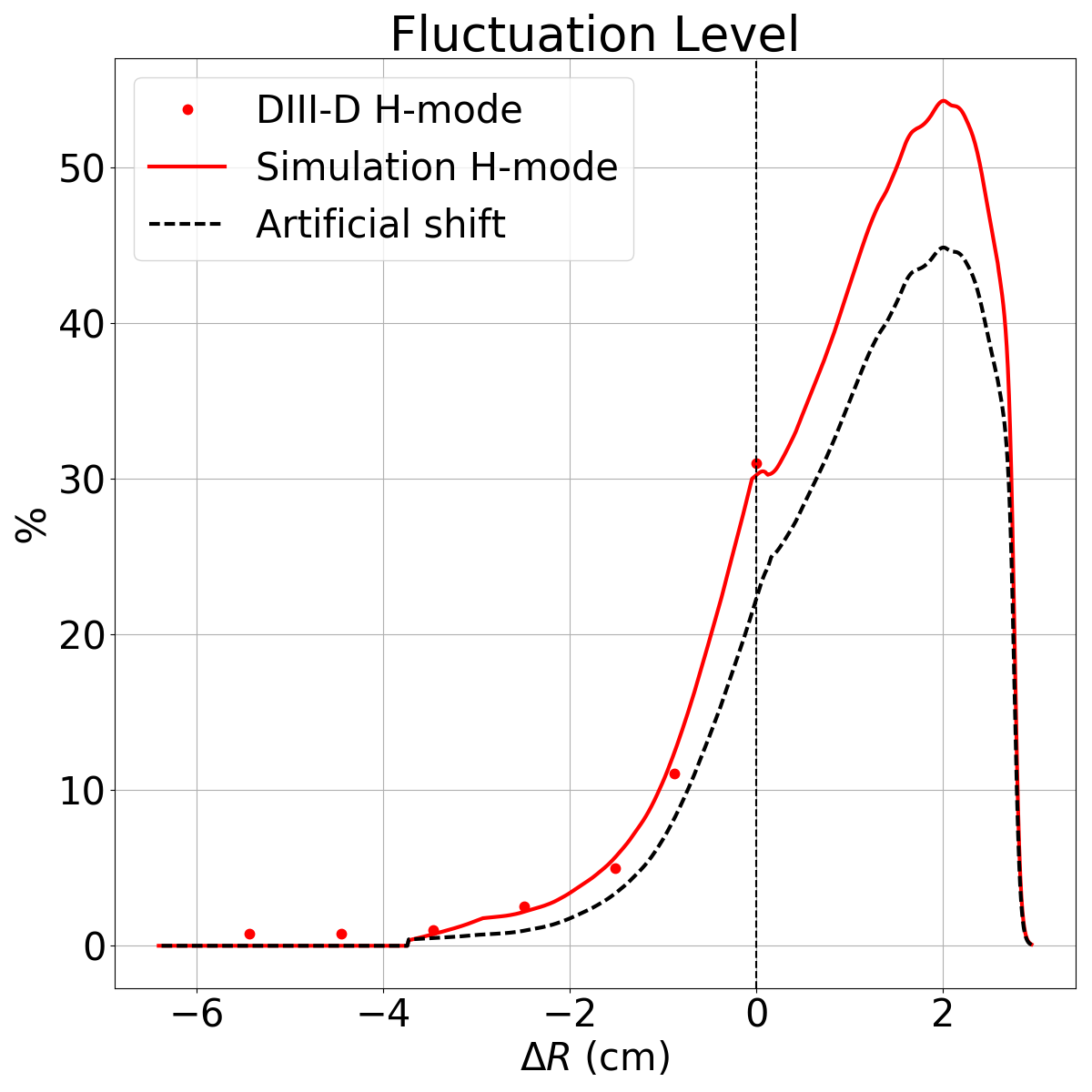}
\end{subfigure}%
\begin{subfigure}{.33\textwidth}
	\includegraphics[width=\linewidth]{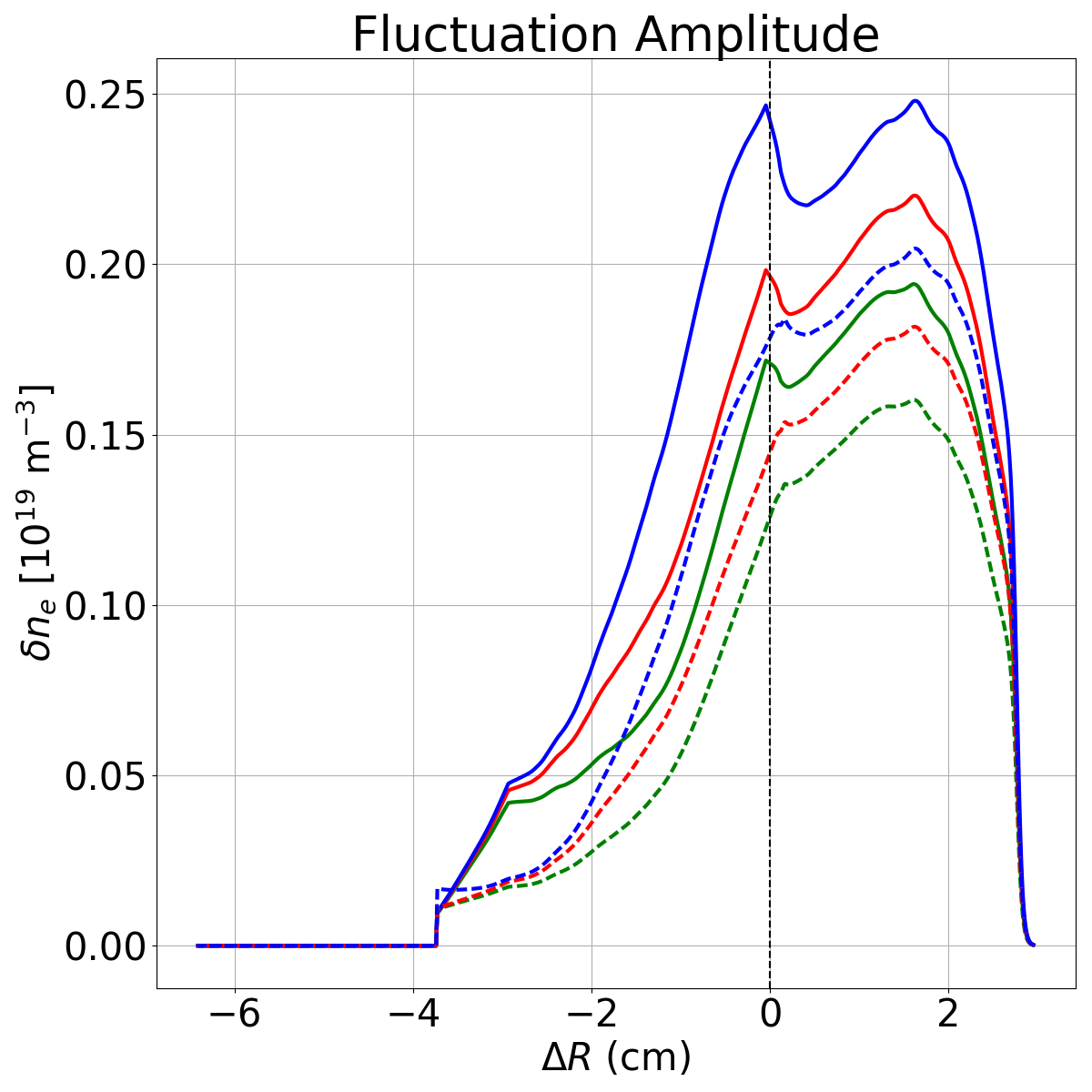}
\end{subfigure}
\caption{The top graph shows the scan in spatial resolution from the nominal location in red. The middle graph shows the scan in fluctuation level error where the red line shows the nominal profile. The bottom graph shows the fluctuation amplitude for the six locations. The colours correspond to the spatial offset where blue represents the density profile closer to the separatrix, red is nominal and green is further away. The solid lines all have the nominal fluctuation level and the dashed lines are lowered by 20\%.}
\label{Fig:Hmode_Diagnostic_scan}
\end{figure}

Both the Thomson scattering and the BES diagnostic systems have spatial resolution of 5 mm. This means a misalignment of the profiles of up to 1 cm is possible. Furthermore the BES has the largest measurement error of up to 20\% in the fluctuation level. Due to the uniquely steep gradients in H-mode of both the fluctuation level profile and the time averaged density profile the combination of these two uncertainties on the calculated beam broadening may be substantial. Modelling provides the ability to artificially scan through these uncertainties to quantify the effect on the predicted broadening. Figure \ref{Fig:Hmode_Diagnostic_scan} shows the six simulations within the diagnostic uncertainty space that have been conducted. The scan through spatial resolution is achieved by moving the time averaged density profile by $\pm5$ mm about the nominal location. For each of these three locations a simulation is run with the nominal fluctuation level and a subsequent simulation with the profile lowered by 20\%.

Table \ref{Tab:H-mode_diagnostic} shows the result for the diagnostic uncertainty. The value used in the broadening prediction is the nominal profile given by the diagnostics corresponding to $\delta r = 0$mm and $\xi_{err}$ = 0. A reduction in the fluctuation level profile by 20\% causes a decrease in the broadening at the absorption surface by 40\%. This is consistent across the scan in spatial resolution. Uncertainty in the spatial location results in an asymmetric change in the calculated broadening. This is expected due to the asymmetric nature of both profiles in space. Moving the density profile by 5 mm results in up to a 50\% change in the background density for the same fluctuation level in the steep density gradient region. Because the background density determines how strongly the wave interacts with the plasma a large change in the scattering is encountered. This results in a change in the predicted broadening at the absorption surface of $+30\%$ for movement towards the separatrix (towards larger fluctuations) and $-40\%$ for movement away (towards smaller fluctuations).

\begin{table}[H]
\small
\begin{center}
    \begin{tabular}{ p{3.2 cm} | p{2.8 cm} | p{2.8 cm}| p{2.8 cm} } \hline
        & $\delta r = +5$mm & $\delta r = 0$mm & $\delta r = -5$mm \\ 
         \hline\hline
     
	$\xi_{err}$ = $0$\% & $\times1.662 \pm 0.132$ & $\times1.515 \pm 0.102$ & $\times1.292 \pm 0.058$ \\
	$\xi_{err}$ = $-20$\% & $\times1.403 \pm 0.080$ & $\times1.302 \pm 0.060$ & $\times1.170 \pm 0.034$ \\
	
    \end{tabular}
\end{center}
\caption{Results for the scan in diagnostic uncertainty. $\xi_{err}$ is the error in fluctuation level and $\delta r$ corresponds to a spatial shift relative to the nominal location. Positive $\delta r$ is a movement towards the separatrix from the nominal and negative the opposite.} 
\label{Tab:H-mode_diagnostic}
\end{table}

\section{Discussion}
\label{Sec:Discus}

By combining experimental measurements with numerical modelling we have been able to show that the main source of ECRH deposition profile broadening is due to scattering of the beam by density fluctuations occurring solely in the edge region in DIII-D. Though fast particle and diffusive effects will occur they should not contribute more than 10\% which is in agreement with previous work \cite{Brookman2017, Brookman2017b, Decker2012, Casson2015}. Furthermore these diffusive effects contribute even less when the beam is already broadened \cite{Poli2015} as is the case here. However, a consistent under-prediction of the broadening of $\approx$10\% is observed in the simulations which is believed to be accounted for by the combination of two sources. Firstly, the fast electron transport is not decoupled from the experimental measurements of the deposition broadening. Secondly, the method for projecting to the absorption region is expected to under-predict the broadening. Both are sources of consistent mismatch between experiment and simulation neither of which are expected to contribute more than 10\% and are contained within the respective errors. 
We have demonstrated the difficulty in modelling H-mode stems from the resolution of the diagnostics. A large change in the calculated beam broadening can occur within the resolution of the diagnostics which accounts for the some what diminished agreement found for this scenario. Doubling of the resolution and a reduction of the measurement error to 10\% from 20\% would be needed to allow for consistent modelling of H-mode. Yet despite the difficulties in making these measurements and matching the simulated plasma profiles to experiment, remarkable agreement with experiment is found across vastly different operating scenarios. It is worth noting that the broadening shown here is only maximised as a negative effect when the normalised poloidal flux, $\psi_N$, is perpendicular to the absorption surface because the broadening means that power is spread across multiple $\psi_N$ surfaces (see figure \ref{Fig:Simulation_domain_in_experiments}). However if the absorption surface is aligned parallel then the fast parallel transport along field lines will negate the broadening (although it is uncertain how practical this may be in experiment).

\vspace{5mm}

Previously beam tracers with a statistical description of the turbulence have been used to calculate the beam broadening for experiments and to predict for ITER \cite{Decker2012, Tsironis2009, Bertelli2010, Peysson2011, Sysoeva2015, Peysson2012}. However it is not yet clear whether such methods capture the full broadening of the wave. Due to the underlying formalism of the wave mechanics, contributions to the scattering from turbulence structures whose characteristic size is comparable to the wavelength will be missed because these lie outside of the beam tracing code's region of validity. Moreover, this has been shown to be the precise region where the scattering is maximised \cite{Kohn2016, Kohn2017}. This is of concern for ITER as the beam broadening, which has already bean calculated as 100\% in these previous studies, could be an underestimation. It should be noted that as explained in section \ref{Sec:Intro} this region of 100\%-200\% broadening is in the limit of operation for fully removing saturated NTM's even with the use of modulation. This is not to say that such a scenario will present itself only that such codes used to predict the broadening for ITER should be benchmarked on a shot by shot basis to experiment and where also possible, the full wave description for the specific benchmark shot. This will be the subject of the future work which will also consider the requirements for robust ITER beam broadening predictions and the potential for reducing the beam broadening and may be possible without significantly affecting the ITER ECCD design. Additionally, within this work the main contributions to the errors on the modelling which come from the method used to project to the absorption surface will be removed. This will be done by using a ray tracer to trace the beam envelope through the experimentally measured density and magnetic field profiles for the core. Further reduction may be possible by matching the end of the simulation domain to a beam tracer, though this has not been considered in detail.

\section{Acknowledgements}
The authors acknowledge access to the EUROfusion High Performance Computer (Marconi-Fusion) through EUROfusion. We also acknowledge access to the HELIOS supercomputer system at Computational Simulation Centre of International Fusion Energy Research Centre (IFERC-CSC), Aomori, Japan, under the Broader Approach collaboration between Euratom and Japan, implemented by Fusion for Energy and JAEA. This material is based upon work supported by the U.S. Department of Energy, Office of Science, Office of Fusion Energy Sciences, using the DIII-D National Fusion Facility, a DOE Office of Science user facility, under Awards DE-FC02-04ER54698. DIII-D data shown in this paper can be obtained in digital format by following the links at https://fusion.gat.com/global/D3D\texttt{\char`_}DMP. One of the authors (MBT) was funded by the EPSRC Centre for Doctoral Training in Science and Technology of Fusion Energy grant EP/L01663X. The authors would like to thank Dr B.D. Dudson, Dr D. Dickinson and Dr D. Hatch for useful discussions.

\end{document}